\documentclass[10pt,fleqn]{article}

\usepackage{latexsym}
\usepackage{amsbsy}
\usepackage{amsfonts}
\usepackage{amssymb}
\usepackage{amscd}
\usepackage{t1enc}
\usepackage{epsfig}
\usepackage{color}
\usepackage{cite}

\usepackage{capt-of}
\usepackage{caption}

\mathchardef\Gamma="0100
\mathchardef\Theta="0102
\mathchardef\Lambda="0103
\mathchardef\Xi="0104
\mathchardef\Pi="0105
\mathchardef\Sigma="0106
\mathchardef\Upsilon="0107
\mathchardef\Phi="0108
\mathchardef\Psi="0109
\mathchardef\Omega="010A

\mathchardef\Delta="7001
\newcommand{\sq}{\hbox{\rlap{$\sqcap$}$\sqcup$}}
\newcommand{\qed}{\ifmmode\sq\else{\unskip\nobreak\hfil
  \penalty50\hskip1em\null\nobreak\hfil\sq
  \parfillskip=0pt\finalhyphendemerits=0\endgraf}\fi{}}
\def\I{{\rm i}}
\def\D{{\rm d}}
\def\E{{\rm e}}

\def\basind#1{{\mbox{$_{#1}$}}}

\def\cal{\mathcal}

\begin{document}

\title{Double-Conjugation Law in  Geometrical Coherent Imaging. \\
Application to Gaussian Beams}

\date{}

\maketitle

\begin{center}

\vskip -.5cm 

{
\renewcommand{\thefootnote}{}
 {\bf   Pierre Pellat-Finet\footnote{\hskip -.53cm Laboratoire de Math\'ematiques de Bretagne Atlantique UMR CNRS 6205,

\noindent Universit\'e de Bretagne Sud, CS 60573, 56017 Vannes, France.

\noindent pierre.pellat-finet@univ-ubs.fr     %\hfill \today
 }}
}
\setcounter{footnote}{0}

\medskip
{\sl \small Univ Bretagne Sud,   CNRS UMR 6205, LMBA, F-56000 Vannes, France}
 
\end{center}

\vskip 1cm
%*************************************ABSTRACT

\begin{center}
\begin{minipage}{12cm}
\hrulefill

\smallskip
{\small
  {\bf Abstract.} A centred system forms the coherent image of the optical field on a sphe\-ri\-cal cap, taken as an object,  on another  spherical cap, whose vertex  and curvature center are the respective paraxial images of the  vertex and center of the object cap. That ``double-conjugation'' law, usually  obtained in the framework of a scalar theory of diffraction, is  deduced from concepts of geometrical optics. A magnification law between radii of conjugate spherical-caps  extends the notion of longitudinal ma\-gni\-fication to finite distances, and generalizes the longitudinal Lagrange--Helmholtz formula. Applying the double-conjugation law  to imaging Gaussian beams shows that those beams  perfectly obey  geometrical optics laws. 


\smallskip
\noindent {\bf Content}%***************************************

\smallskip

\noindent 1. Introduction \dotfill \pageref{sect1}

\noindent 2. The laws of coherent imaging \dotfill \pageref{sect2}

\noindent 3. Interpreting the radius-magnification law  \dotfill \pageref{sect3}

\noindent 4. Application to Gaussian beams \dotfill \pageref{sect4}

\noindent 5. Conclusion\dotfill \pageref{conc}

\noindent  References\dotfill \pageref{ref}

}

\hrulefill
\end{minipage}
\end{center}

\bigskip

%**********************************
\section{Introduction}\label{sect1}
%**********************************

\subsection{Necessity of regarding imaging from the view of coherent optics}\label{sect11}%*********************************

A great part of geometrical optics deals with image formation by centred systems. Paraxial geo\-metrical optics, or Gaussian optics,  constitutes a first-order approximation of image formation: angles of light rays with the optical axis of a centered system and  transverse dimensions of  objects and images are small enough that in actual calculations we may replace, for example, $\sin \alpha$ and $\tan\alpha$ by $\alpha$, $ \cos\alpha$ by 1, etc. The considered objects and their images generally lie in planes, orthogonal to the optical axis.
Expansions to higher orders are necessary to deal with geometrical aberrations: they are at least of 3rd  order \cite{Mar1,Bor}.

Moreover, geometrical optics essentially deals with incoherent objects and is therefore only interested in the irradiance of an image and not in the corresponding optical-field amplitude. An irradiance is a quadratic magnitude, proportional to the square modulus of the optical-field amplitude: thus,  irradiances on a spherical cap and on the plane tangent to the sphere at its vertex are equal, %within the limits of paraxial approximation,
since the field amplitudes on each surface differ by a quadratic phase factor.
In other words, and as long as we are not concerned with geometrical aberrations,  phases of optical fields, which do not appear in  irradiances, have not to be taken into account in paraxial geometrical optics. Approximating  object surfaces by planes is consistent with negelecting phase differences due to surface profiles.

Since the advent of lasers in 1960, however, coherent optics has taken on new importance. Some applications require controlling phases on images. For example in optical signal processing (VanderLugt filters and correlators \cite{Van,Goo1,PPF1}) field amplitudes are projected on planar holograms---used as filters---and their phases must exactly match  phases engraved in those holograms.
Given that an image is a copy of the considered object, the issue arises under what conditions  the phase on the object is preserved in the imaging.   It is necessary then to study the image of an object both in amplitude and phase,  that is, to study the coherent image of a given object.
We will show that the coherent image of a planar object generally lies on a spherical cap. This is why we shall use spherical caps, more general than planar objects and images (planes will be regarded as spheres with infinite curvature radii).  Since the phase difference between the field on a spherical cap and the field on a tangent plane is a quadratic phase factor, a first-order approximation is not enough to deal with coherent images; we have then to develop a second-order approximation of the field amplitude and of its transfer through a lens, that is, a ``metaxial approximation'' \cite{PPF1,GB1,GB2}.

The issue of coherent imaging is usually analyzed in the framework of a scalar theory of diffraction \cite{PPF1,GB1,GB2,Kog}. In the present article we establish the ``double-conjugation law'' of coherent imaging, and some of related results,  within the framework of geometrical optics, without recourse to the theory of diffraction (nor to physical optics).  The developed theory is purely geometrical in the sense that limited apertures of optical systems are not taken into account (i.e. diffraction by pupils are not taken into account). We eventually apply the previous law to Gaussian beams and show that  imaging them through objective lenses can adequately be managed in the framework of geometrical optics.

%*********************************************************
\subsection{Coherent object, coherent image}\label{sect12}
%*********************************************************

A luminous object is coherent if the vibrations emitted by all its various points  have between themselves permanent phase relationships over time. That is spatial coherence.

In the following, for reasons given in Section \ref{sect11}, objects and images are spherical caps having in general an axial symmetry around the optical axis of the centered system under study (planes are included, as mentioned above). As always in optics, such spherical caps may be real or virtual. 
Lightwaves are assumed to be quasi-monochromatic (wavelength $\lambda$).

The coherent image of a coherent object is a homothetic copy of the object in amplitude and phase. The scale ratio will be shown to be the lateral magnification, as defined in paraxial optics. This means that we have a coherent image of an object if:
\begin{itemize}
\item The amplitude of the field at a point of the image is equal to the amplitude of the field at the corresponding object-point.
\item The phase difference between vibtations at two points of the image is equal to the phase difference between vibrations at the corresponding  points of the object (``preservation'' of the phase in imaging).

\end{itemize}
Coherent imaging between an object and its image will be expressed by Eq.\ (\ref{eq33}). %(\ref{eq4}).

%**************************************************
\section{The laws of coherent imaging}\label{sect2}
%**************************************************

\subsection{Paraxial formulae}

We recall some usual formulae of paraxial geometrical optics.
We adopt the convention that algebraic measures are positive if taken in the sense of light propa\-gation. The rule also holds for mirrors, which means that the positive sense is not the same before and after reflection; the advantage of doing this way is that the following formulae hold true for dioptric as well as for katoptric systems.

Newton's conjugation formula for a centered system with foci $F$ (object focus) and $F'$ (image focus)  is
\begin{equation}
  z\,z'=f\, f'\,,\end{equation}
where $f$ is the object focal-length, $f'$ the image focal-length, and $z=\overline{FA}$, $z'=\overline{F'A'}$, point $A$ being on the optical axis and $A'$ being its paraxial image (also on the optical axis). Points $A$ and $A'$ are conjugates. We have $f'/n'=-f/n$, where $n$ is the refractive index of the object space, and $n'$ of the image space.

If $\overline{AB}$ is a transverse object  and $\overline{A'B'}$ its paraxial image, the lateral magnification for that conjugation is defined by
\begin{equation}
  m={\overline{A'B'}\over\overline{AB}}\,,\end{equation}
and if $A$ and $A'$ are on the optical axis, then ($z$ and $z'$ are defined above)
\begin{equation}
  m=-{f\over z}=-{z'\over f'}\,.\label{eq3}\end{equation}

If $H$ and $H'$ are the principal points of the system (on the optical axis), we denote $p=\overline{HA}$ and $p'=\overline{H'A'}$, and the Descartes's conjugation formula (with origin at principal points) is
\begin{equation}
  {n'\over p'}={n\over p}+{n'\over f'}\,,\end{equation}
the lateral magnification $m$ (defined above) being such that
\begin{equation}
  m={n \, p'\over n'p}\,.\label{eq5}\end{equation}

For a refracting spherical surface (or spherical cap)  whose vertex is $V$, center is $C$ and curvature radius is $R=\overline{VC}$, we have $f'=n'R/(n'-n)$, so that the Descartes's formula becomes
\begin{equation}
  {n'\over p'}={n\over p}+{n'-n\over R}\,,\label{eq6a}\end{equation}
where $p=\overline{VA}$ and $p'=\overline{VA'}$ (because $H\equiv V\equiv H'$).

\subsection{Phase difference between two tangent spherical caps}%*****************************

\begin{figure}%$$$$$$$$$$$$$$$$$$$$$$$$$$$$$$$$$$$$$$$$$$$$$$$
  \begin{center}
  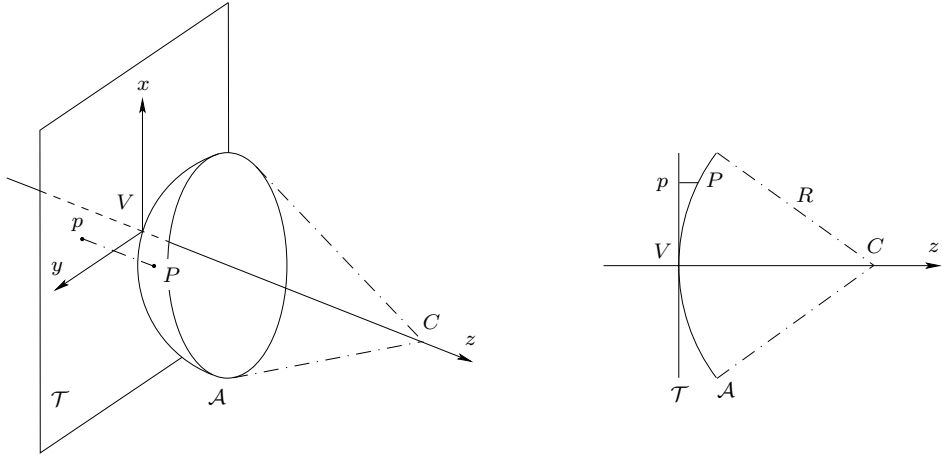
  \caption{\small Coordinates $x$ and $y$ of $p$ in the plane ${\cal T}$ are also those of $P$ on the spherical cap ${\cal A}$.\label{fig1}}
  \end{center}
\end{figure}

Consider a spherical object, in fact a spherical cap ${\cal A}$ with vertex $V$ and center of curvature $C$ (Fig.\ \ref{fig1}). The radius of the sphere is $R=\overline{VC}$. Let ${\cal T}$ be the  plane tangent to the sphere  at $V$. The object is assumed to be coherent.

%$$$$$$$$$$$$$$$$$$$$$$$$$$$$$$$$$$$$$$$$$$$$$$$$$

If $P$ is a point of the sphere and $p$ its projection on ${\cal T}$,   and if we relate the plane ${\cal T}$ to axes $Vx$ and $Vy$ (Fig.\ \ref{fig1}), the coordinates $x_p$ and $y_p$ of $p$ in ${\cal T}$ can serve as coordinates  of $P$ on the sphere ${\cal A}$.  (There is no ambiguity in defining a point $P$ of the spherical cap ${\cal A}$ by $x_p$ and $y_p$, as long as ${\cal A}$ is less than a half-sphere.) If $U_A(P)$ denotes the amplitude of the optical field on ${\cal A}$ at point $P$, and $U_T(p)$ the amplitude on ${\cal T}$ at point $p$, then, up to second order in $x_p$ and $y_p$,
\begin{equation}
   U_A(P)=U_A(x_p,y_p)=U_T(x_p,y_p)\,\exp\left[-{\I\pi \over \lambda R}(x_p^2+y_p^2)\right]\, .\label{eq1}\end{equation}
Thereafter we will write no more subscripts and will denote $x$ and $y$ the coordinates of a point on the spherical cap ${\cal A}$.

\begin{figure}[h]%$$$$$$$$$$$$$$$$$$$$$$$$$$$$
  \begin{center}
  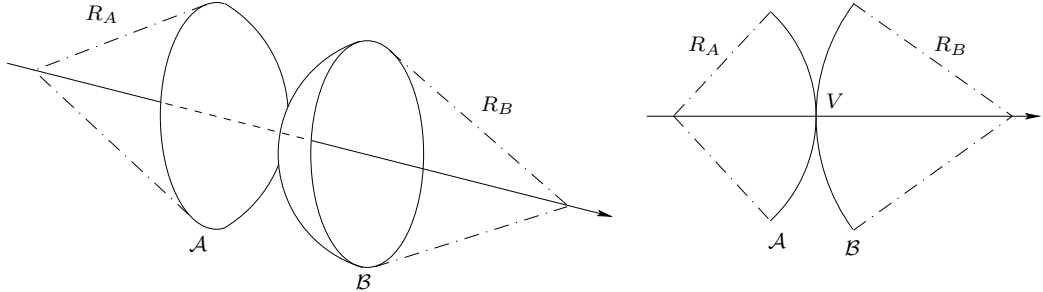
  \caption{\small Field  amplitudes on the tangent spherical caps ${\cal A}$ and ${\cal B}$ differ by a quadratic phase factor (up to second order), see Eq.\ (\ref{eq2}). \label{fig2}}
  \end{center}
\end{figure}

Now consider two spherical caps ${\cal A}$ and ${\cal B}$ tangent at their common vertex $V\!$, with respective radii of curvature $R_A$ and $R_B$ (Fig.\ \ref{fig2}). By considering the tangent plane common to the two spheres as intermediate, we deduce from Eq.\ (\ref{eq1})
\begin{equation}
   U_B(x,y)=U_A(x,y)\,\exp \left[-{\I\pi \over\lambda}\left({1\over R_B}-{1\over R_A}\right) (x^2+y^2)\right]\,.\label{eq2}\end{equation}
Equation (\ref{eq2}) shows that there is a quadratic phase-difference between the amplitudes of the fields on two tangent spheres, or even between the amplitudes of the fields on a plane and on a sphere. If we are dealing with coherent imaging, i.e. if we wish to obtain an image in amplitude and phase, it is necessary to distinguish the field amplitude on a spherical cap from that on the tangent plane at vertex, as already mentioned.

%$$$$$$$$$$$$$$$$$$$$$$$$$$$$$$$$$

\subsection{Imaging a spherical cap concentric to a refracting spherical cap}\label{sect23}%***********

We consider a refracting spherical cap  ${\cal D}$ with vertex $V$ and curvature center $C$ (radius $R=\overline{VC}$), separating  two isotropic and homogeneous media with respective indices $n$ (object space) and $n'$ (image space).

The analysis is carried out in the conditions of approximate stigmatism. Let $A$ be  a point on the axis of ${\cal D}$: its paraxial image  is  $A'\!$, whose  position  is given, on the axis, by the conjugation formula of the refracting spherical cap, see Eq.\ (\ref{eq6a}).

\begin{figure}%[h]%$$$$$$$$$$$$$$$$$$$$$$$$$$$$$$$$$$$$$$$
  \begin{center}
  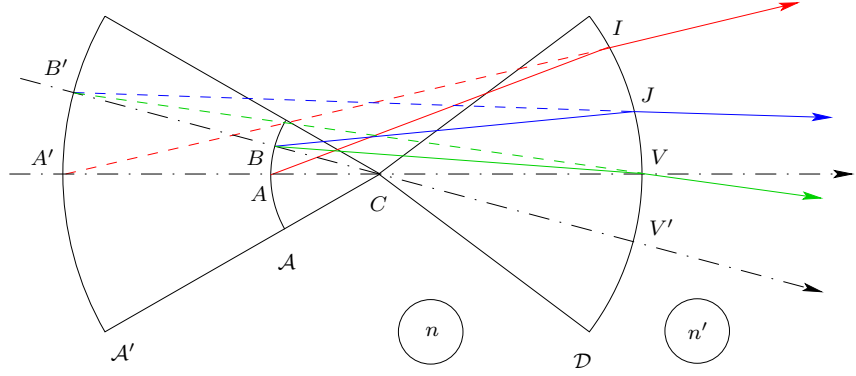
  \caption{\small Imaging a spherical cap ${\cal A}$, concentric to the refracting spherical cap ${\cal D}$. If  points $A$ and $A'$ are conjugates,  points $B$ and $B'$, deduced from $A$ and $A'$ by rotation around $C$, the center of curvature of ${ \cal D}$, also are conjugates, and the optical paths $[AIA']$ and $[BJB']$ are equal. Here ${\cal A}$ is a real object and ${\cal A}'$ a virtual image.\label{fig3}}
  \end{center}
\end{figure}

\begin{figure}[b]%$$$$$$$$$$$$$$$$$$$$$$$$$$$$$$$$$$$$$$$$
  \begin{center}
  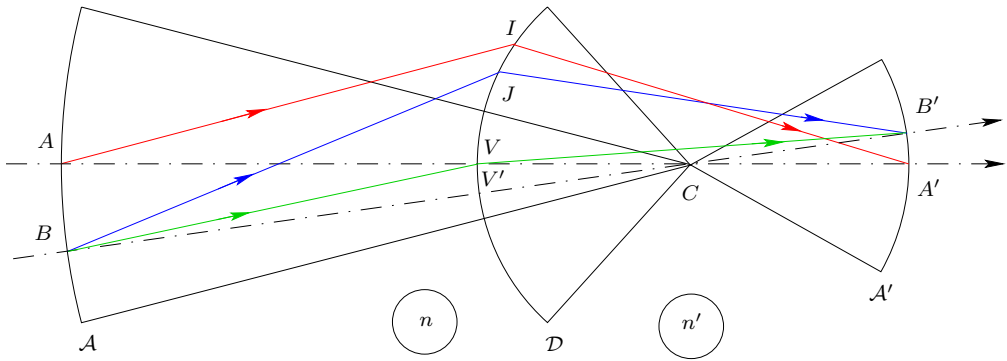
  \caption{\small Same situation as in Fig.\ \ref{fig3}, but for  a real object ${\cal A}$ and a real image ${\cal A}'$. \label{fig4}}
  \end{center}
\end{figure}

Let us rotate the figure around $C$ (Figs.\ \ref{fig3} and \ref{fig4}). The vertex of ${\cal D}$  comes in $V'\!$, point $A$ in $B$ and point $A'$ in $B'\!$. The sphere ${\cal D}$ being globally invariant in the rotation, the point $B'$ is the image of $B$ by ${\cal D}$. Let $I$ be a point on ${\cal D}$, which comes in $J$ in the rotation:  the optical path $[AIA']$ (which contains a virtual part, if drawn as in Fig.\ \ref{fig3}) is transformed into  $[BJB']$, and we have $[BJB']=[AIA']$. According to Fermat's principle \cite{Mar1}, the optical path $[AIA']$ is constant up to second order, whatever the ray issued from $A$, i.e.  whatever  point $I$ on ${\cal D}$  (approximate stigmatism).  The same result holds for $[BJB']$,  whatever  point $J$ on ${\cal D}$, because $B'$ is the image of $B$. It follows that the phase difference between vibrations at $A'$ and $B'$ is equal to the phase difference between vibrations at $A$ and $B$.

%$$$$$$$$$$$$$$$$$$$$$$$$$$$$$$$$$$$$$$$$$$$$$

%$$$$$$$$$$$$$$$$$$$$$$$$$$$$$$$$$$$$$$$$$$$$$

The spherical cap ${\cal A}$ passing through $A$ and centered at $C$ contains  point $B$, and point $B'$ belongs to the sphere ${\cal A}'$ passing through $A'$ and centered at $C$. The mapping: $B\longmapsto B'$ is a homothecy between ${\cal A}$ and ${\cal A}'$. Moreover, for every $B_1$ and every $B_2$ belonging to ${\cal A}$, their images $B'_1$ and $B'_2$ belong to  ${\cal A}'\!$, and the phase difference between vibrations at $B_1'$ and $B_2'$ is equal to the phase difference between vibrations at $B_1$ and $B_2$ (because the phase difference between $B'_j$ and $A'$ is equal to the phase difference between $B_j$ and $A$). This means that ${\cal A}'$ is the coherent image of ${\cal A}$, and field amplitudes
 $U_A$ and $U_{A'}$ are homothetic 
\begin{equation}
  U_{A'}(x',y')={1\over K}\,U_A\!\left({x'\over K},{y'\over K}\right)\,,\label{eq4}\end{equation}
where $K$ is the scale factor.
The factor $1/K$ in front of $U_A$ expresses the conservation of energy (or power): indeed, we must have
\begin{equation}
  \int_{{\cal A}'} | U_{A'}(x',y')|^2\,\D x'\,\D y'= \int_{\cal A} | U_{A}(x,y)|^2\,\D x\,\D y\,.\end{equation}

Since ${\cal A}$ and ${\cal A}'$ are homothetic, the scale factor  $K$ such that
\begin{equation}
  K={R_{A'}\over R_A}\,,\label{eq11}\end{equation}
where $R_A$ is the radius of ${\cal A}$, and $R_{A'}$ the radius of ${\cal A}'$ ($R_A=\overline{AC}$ and $R_{A'}=\overline{A'C'}$, in Figs.\ \ref{fig3} and \ref{fig4}).
The ray $BV$ (drawn in green in Figs.\ \ref{fig3} and \ref{fig4}) is refracted in ray $VB'$ (or $B'V$).  We denote $p=\overline{VA}$ and $p'=\overline{VA'}$, so that  Snell's law at $V$, approximated to first order (Kepler's law \cite{Mar1}), leads to
\begin{equation}
  K={n\,p'\over n'p}\,.\label{eq12}\end{equation}
that is, to $K=m$, according to Eq.\ (\ref{eq5}). The scale factor between the two homothetic caps ${\cal A}$ and ${\cal A}'$ is equal to the  lateral magnification for the conjugation of $A$ and $A'$.

Equation (\ref{eq4}) then becomes
\begin{equation}
  U_{A'}(x',y')={1\over m}\;U_A\!\left({x'\over m},{y'\over m}\right)\,.\label{eq4b}\end{equation}

Fresnel transmission coefficients of the refracting surface are not taken into account. Introducing them does not change imaging laws. In Eq.\ (\ref{eq4}) we omit a phase factor of the form $\exp (-2\I\pi \nu \tau)$, where $\tau$ denotes the time for light to travel from $A$ to $A'$ ($\nu$ denotes lightwave frequency).  This factor is common for all points of ${\cal A}$ and their images and does not affect coherent imaging.

\medskip\noindent {\bf Remark.}  The above is developed within the framework of approximate stigmatism. However, if $A$ and $A'$ are the Weierstrass (or Young) points of the refracting spherical cap ${\cal D}$, for which there is rigorous stigmatism, there is still rigorous stigmatism for points $B$ and $B'\!$, deduced from $A$ and $A'$ by rotation around the center of ${\cal D}$ (Figs.\ \ref{fig3} and \ref{fig4}). Finally, there is  rigorous stigmatism between every point of the spherical cap ${\cal A}$ and its image on ${\cal A}'$. It is a case of rigorous stigmatism between spheres. This somehow generalizes the aplanatism condition of classical geometrical optics (Weierstrass points are aplanatic).

\subsection{Imaging an arbitrary spherical cap}%*****************

\begin{figure}%[b]%$$$$$$$$$$$$$$$$$$$$$$$$$$$$$$$$$$$$$$$
  \begin{center}
  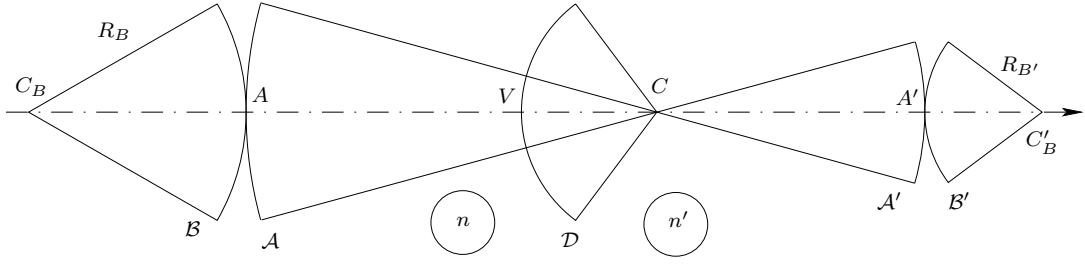
  \caption{\small Imaging a spherical cap ${\cal B}$, not concentric to ${\cal D}$. \label{fig4b}}
  \end{center}
\end{figure}

We consider a spherical cap ${\cal A}$, concentric to ${\cal D}$, and its coherent image ${\cal A}'$ as in Section \ref{sect23}.  We then consider  a spherical cap   ${\cal B}$ tangent to  ${\cal A}$ and centred at $C_B$ (radius $R_B=\overline{AC_B}$), see Fig.\ \ref{fig4b}. We denote $U_B$ the field amplitude on ${\cal B}$. We have
\begin{equation}
  U_A(x,y)=U_B(x,y)\,\exp \left[-{\I\pi\over \lambda}\left({1\over R_A}-{1\over R_B}\right)(x^2+y^2)\right]\,.\end{equation}
The field amplitude on  ${\cal A}'$ is given by Eq.\  (\ref{eq4}).

Let  ${\cal B}'$ be a spherical cap tangent to  ${\cal A}'$ and  centred at $C_{B'}$ (radius $R_{B'}$).  Since  ${\cal B}'$ lies in the image space, the field amplitude on  ${\cal B}'$ is written
\begin{eqnarray}
  U_{B'}(x',y')\!\!\!\!&=&\!\!\!\!U_{A'}(x',y')\,\exp\left[-{\I\pi \over \lambda '}\left({1\over R_{B'}}-{1\over R_{A'}}\right) (x'^2+y'^2)\right]\nonumber \\
  &=& \!\!\!\! {1\over m}U_B\left({x'\over m},{y'\over m}\right)
  \exp \left[-{\I\pi\over \lambda}\left({1\over R_A}-{1\over R_B}\right)\left({x'^2+y'^2\over m^2}\right)\right]\nonumber \\
  & & \hskip 3cm \times \;\;\;
  \,\exp\left[-{\I\pi \over \lambda '}\left({1\over R_{B'}}-{1\over R_{A'}}\right) (x'^2+y'^2)\right]
  \,,\label{eq14}\end{eqnarray}
where $\lambda '$ is the wavelength in the image space ($n'\lambda '=n\lambda$).

%$$$$$$$$$$$$$$$$$$$$$$$$$$$$$$$$$$$$$$$$$$$$$

The field on ${\cal B}'$ is the coherent image of the field on ${\cal B}$ if, and only if, the arguments of the two exponential are opposite for every $x'$ and $y'$, that is, if, and only if, 
\begin{equation}
  {1\over m^2\lambda}\left({1\over R_A}-{1\over R_B}\right)=
  {1\over \lambda '}\left({1\over R_{A'}}-{1\over R_{B'}}\right)\,.\label{eq9}\end{equation}
We denote $q=\overline{VC_B}$ and $q'=\overline{VC_{B '}}$. (We still denote $p=\overline{VA}$ and $p'=\overline{VA'}$.) Then
  \begin{equation}
    R_B=\overline{AC_B}=q-p\,,\hskip 1cm R_{B'}=\overline{A'C_{B'}}=q'-p'\,.\end{equation}
We also have ($R$ is the curvature radius of ${\cal D}$)
  \begin{equation}
    R_A=\overline{AC}=R-p\,,\hskip 1cm R_{A'}=\overline{A'C}=R-p'\,,\end{equation}
 and since $n\lambda =n'\lambda '$, and taking Eq.\  (\ref{eq5}) into account, we  write Eq.\  (\ref{eq9}) in the form 
  \begin{equation}
    {p^2\over n}\left({1\over R-p}-{1\over q-p}\right)
    ={p'^2\over n'}\left({1\over R-p'}-{1\over q'-p'}\right)\,,\label{eq19}\end{equation}
that is
  \begin{equation}
    {p^2\over n}\;{q-R\over (R-p)(q-p)}={p'^2\over n'}\;{q'-R\over (R-p')(q'-p')}\,.\label{eq20}\end{equation}
Equations  (\ref{eq11}) and (\ref{eq12}) give
  \begin{equation}
    {n\,p'\over n' p}={R_{A'}\over R_A}={R-p'\over R-p}\,,\end{equation}
from which we deduce
  \begin{equation}
    {p\over n(R-p)}={p'\over n'(R-p')}\,,\end{equation}
 so that Eq.\   (\ref{eq20}) becomes
  \begin{equation}
    {p(q-R)\over q-p}= {p'(q'-R)\over q'-p'}\,.\label{eq17}\end{equation}
 We divide the numerator and the denominator of the left side of Eq.\ (\ref{eq17}) by $pq$, and those of the right side by $p'q'$, and we divide both numerators by $R$: we obtain 
  \begin{equation}
    {\displaystyle{1\over R}-{1\over q}\over \displaystyle{1\over p}-{1\over q}}
    ={\displaystyle{1\over R}-{1\over q'}\over \displaystyle{1\over p'}-{1\over q'}}\,,
  \end{equation}
which can also be written 
  \begin{equation}
    {\displaystyle{n\over R}-{n\over q}\over \displaystyle{n\over p}-{n\over q}}
    ={\displaystyle{n'\over R}-{n'\over q'}\over \displaystyle{n'\over p'}-{n'\over q'}}\,.
  \end{equation}
 We eventually substract the denominators from the numerators to obtain
   \begin{equation}
    {\displaystyle{n\over R}-{n\over p}\over \displaystyle{n\over p}-{n\over q}}
    ={\displaystyle{n'\over R}-{n'\over p'}\over \displaystyle{n'\over p'}-{n'\over q'}}\,.
    \label{eq26}
   \end{equation}
   According to the conjugation formula  (\ref{eq6a}), the numerators in Eq.\ (\ref{eq26}) are equal to each other and necessarily the denominators must be equal to each other, that is
   \begin{equation}
     {n\over p}-{n\over q}= {n'\over p'}-{n'\over q'}\,.\label{eq21}\end{equation}
Then
   \begin{equation}
     {n'\over q'}-{n\over q}={n'\over p'}-{n\over p}={n'-n\over R}\,,\end{equation}
that is
   \begin{equation}
     {n'\over q'}={n\over q}+{n'-n\over R}\,.\label{eq23}\end{equation}
   Equation (\ref{eq23}) is no more than Eq.\ (\ref{eq6a}) applied to curvature centers $C_B$ and $C_{B'}$ of  ${\cal B}$ and ${\cal B}'$ (that is, changing $p$ into $q$ and $p'$ into $q'$): it means that $C_B$ and $C_{B'}$ are conjugates.

   We conclude: the field on ${\cal B}'$ is the coherent image of the field on  ${\cal B}$ if, and only if, the curvature center of  ${\cal B}'$ is the (paraxial) image of the curvature center of  ${\cal B}$.

   Let us express the previous result in a general form, that is, considering ${\cal B}$ first. Thus, let ${\cal B}$ be  an arbitrary  spherical cap, with vertex $V_B$ and center of curvature $C_B$. To find the coherent image of the field on this sphere, we introduce the sphere ${\cal A}$ tangent to ${\cal B}$ (it therefore passes through $V_B$) and centered at the center of curvature of ${\cal D}$. The coherent image of ${\cal A}$ is the spherical cap ${\cal A}'$ passing through $V_{B'}$, the paraxial image of $V_B$, and concentric to ${\cal D}$. The previous reasoning then shows that the coherent image of ${\cal B}$ is the sphere ${\cal B}'$ tangent to ${\cal A}'$ (it passes through $V_{B'}$) and centered at $C_{B'}$, the paraxial image of $C_B$.

    If $m$ denotes the lateral magnification between vertices $V_B$ and $V_{B'}$, the field amplitudes on  ${\cal B}'$ and  ${\cal B}$ are related by the following equation
   \begin{equation}
     U_{B'}(x',y')={1\over m}\;U_B\!\left({x'\over m},{y'\over m}\right)\,,\label{eq29}\end{equation}
   up to a constant phase factor that depends on the propagation time from $V_B$ to $V_{B'}$. (Equation (\ref{eq29}) is Eq.\ (\ref{eq14}) in which quadratic phase factors cancel, since $C_B$ and $C_{B'}$ are conjugates.)

   \bigskip
   \noindent{\bf Proposition (Double conjugation for a refracting spherical-cap).}
{\it A refracting spherical-cap  forms the coherent image of a spherical object  on the spherical cap whose vertex and center of curvature are the respective paraxial images of the vertex and curvature center of the object spherical-cap.}

\bigskip

   \subsection{Radius-magnification formula for a refracting spherical-cap}%************************************
 We first adapt notations. In the imaging of a spherical cap, we denote $m_{\rm v}$ the lateral magnification at  vertices (it is the previous parameter $m$). And we denote $m_{\rm c}$ the lateral magnification at  centers of curvature (which are conjugates). If $p$ and $p'$ are the algebraic measures of the vertices, with origin at the vertex of ${\cal D}$, and if $q$ and $q'$ are the algebraic measures of the centers of curvature, according to the laws of paraxial optics we have
    \begin{equation}
      m_{\rm v}={n\,p'\over n'p}\,,\hskip 1cm {\rm and}\hskip 1cm m_{\rm c}={n\,q'\over n'q}\,.\end{equation}

    If $R_B$ denotes the radius of the object spherical-cap and  $R_{B'}$ that of the image cap, the radius magnification is defined by 
  \begin{equation}
    m_{\rm r}={R_{B'}\over R_B}\,.\end{equation}

   We use  Eq.\ (\ref{eq21}) to write
  \begin{equation}
    m_{\rm r}={R_{B'}\over R_B}={q'-p' \over q-p}={np'q'\over n'p\,q}={n'\over n}\,m_{\rm v}\,m_{\rm c}\,.\label{eq32}\end{equation}
  Equation (\ref{eq32}) is the formula of radius magnification for a refracting spherical-cap.

  \subsection{Coherent imaging by a centered system}%*****************************

  We note first that the method used to deal with coherent imaging through a refracting spherical-cap can be directly  transposed to imaging by a spherical mirror.

A centered system being a sequence of refracting spheres or spherical mirrors, the property of double conjugation being in a way a ``transfer'' relation, it should be clear  that this property remains valid for every centered system: every refracting spherical-cap (or mirror) composing the  system images an intermediate coherent-object in accordance with the double-conjugation law. Thus every spherical cap ${\cal A}$ in the object space has a first image ${\cal A}_1$ through the first refracting surface (or mirror) of the centered system, the vertex and the center of curvature of ${\cal A }_1$ being the paraxial images of the vertex and the center of curvature of ${\cal A}$. The second refracting surface images ${\cal A}_1$ into a sphere ${\cal A}_2$ whose vertex and center are the paraxial images of the vertex and the center of curvature of ${\cal A}_1$, and which are also the paraxial images of the vertex and the center of curvature of ${\cal A}$ through the subsystem made up of the first two refracting surfaces (or mirrors). And so on.
The result is expressed by the following theorem.

\bigskip\noindent{\bf Theorem 1 (Double conjugation for coherent imaging by a centered system). }  {\it A centred system forms the coherent image of a spherical cap ${\cal A}$ on the spherical cap ${\cal A}'$, whose vertex is the paraxial image of the vertex of  ${\cal A}$, and whose center is the paraxial image of the center of  ${\cal A}$. If $m_{\rm v}$ denotes the lateral magnification for the vertex conjugation, the field amplitudes  on ${\cal A}$ and ${\cal A}'$ are linked by
\begin{equation}
        U_{A'}(x',y')={1\over m_{\rm v}}\,U_A\left({x'\over m_{\rm v}},{y'\over m_{\rm v}}\right)\,,\label{eq33}\end{equation}
up to a constant phase factor (which does not depend on $x'$ and $y'$).}

\bigskip
\noindent{\bf Remark.} The conjugation of vertices is no more that what is obtained by paraxial geometrical optics. But the conjugation of curvature centers is properly a condition of coherent optics. It results from preserving phases in the imaging process.

\bigskip

\noindent{\bf Remark.}  As far as we know, the conjugation of curvature centers has first be mentioned  by Kogelnik in an article devoted to optical resonators \cite{Kog}. The double-conjugation law is more systematicaly stated by Bonnet  in his articles on metaxial optics \cite{GB1,GB2}. In all those works the law is deduced from a scalar theory of diffraction.

\bigskip

We now prove the following theorem.

\bigskip\noindent{\bf Theorem 2 (Bonnet's law of radius magnification).} {\it 
  In the coherent imaging from ${\cal A}$ (radius $R_A$) to ${\cal A}'$ (radius $R_{A'}$) by a centered system, if $m_{\rm v}$ is the lateral magnification at vertices and $m_{\rm c}$ the lateral magnification at  centers of curvature, if $n$ is the refractive index of the object space and $n'$ that of the image space, the radius magnification, defined by $m_{\rm r}=R_{A'}/R_A$, is such that}
\begin{equation}
  m_{\rm r}={n'\over n}\,m_{\rm v}\,m_{\rm c}\,.\label{eq31a}\end{equation}

\bigskip\noindent{\it Proof.} (i) Let ${\cal S}$ be a centered system and let ${\cal D}$ be a refracting spherical-cap. Light is assumed to pass across ${\cal S}$ first, so that ${\cal S}$ and ${\cal D}$ form a centered system, denoted ${\cal S}\cup{\cal D}$.  The successive optical spaces are: object space (refractive index $n$), intermediate space (index $N$) and image space (index $n'$). Let $\overline{AB}$ be a transverse object  in the object space, with $A$ on the optical axis. Let $\overline{A_NB_N}$ be its image by ${\cal S}$ (lateral magnification $m_{\basind N}$), and let $\overline{A'B'}$ be the image of $\overline{A_NB_N}$ by ${\cal D}$ (magnification $m'$). Then $\overline{A'B'}$ is the image of $\overline{AB}$ by ${\cal S}\cup{\cal D}$; the corresponding lateral magnification is $m$, with
\begin{equation}
  m={\overline{A'B'}\over\overline{AB}}={\overline{A'B'}\over \overline{A_NB_N}}\cdot {\overline{A_NB_N}\over \overline{AB}}=m'\,m_{\basind N}\,.\label{eq32a}
  \end{equation}

\smallskip
\noindent (ii) We now prove Eq.\ (\ref{eq31a}) by induction. We note that Eq.\ (\ref{eq31a}) holds for a refracting spherical-cap, see Eq.\ (\ref{eq29}).
Let ${\cal S}$ be a centered system made up of $J$ refracting spheres ($J>0$). Let ${\cal A}_N$ (vertex $V_N$, center $C_N$, radius $R_N=\overline{V_NC_N}$) be the coherent image of ${\cal A}$ (vertex $V$, center $C$, radius $R=\overline{VC}$) through ${\cal S}$: the corresponding lateral magnifications are $m_{{\rm r}N}$ at vertices and $m_{{\rm c}N}$ at centers.
Let  ${\cal D}$ be a refracting spherical cap located after ${\cal S}$ and let ${\cal A}'$ be the coherent image of ${\cal A}_N$ through ${\cal D}$:  lateral magnifications are $m'_{\rm r}$ and $m'_{\rm c}$.

We denote by ${\cal S}\cup {\cal D}$ the system made up of ${\cal S}$ followed by ${\cal D}$.
The vertex $V'$ of ${\cal A}'$ is the paraxial image of $V_N$ through ${\cal D}$ and is also the paraxial image of $V$ through ${\cal S}\cup {\cal D}$. Similarly, the center $C'$ of ${\cal A}'$ is the paraxial image of $C_N$ through ${\cal D}$ and is also the paraxial image of $C$ through ${\cal S}\cup {\cal D}$. We conclude that ${\cal A}'$ is the coherent image of ${\cal A}$ through ${\cal S}\cup {\cal D}$. Let $m_{\rm r}$ be the lateral magnification at vertices and $m_{\rm c}$ at centers.
According to item (i), the magnification at vertices is
\begin{equation}
  m_{{\rm v}}=m'_{\rm v}\,m_{{\rm v}N}\,,\end{equation}
and that at  centers is
\begin{equation}
  m_{\rm c}=m'_{\rm c}\,m_{{\rm c}N}\,.\end{equation}
We  deduce
\begin{equation}
  m_{\rm r}={R_{A'}\over R_A}={R_{A'}\over R_N}\cdot{R_N\over R_A}={n'\over N}\,m'_{\rm v}\,m'_{\rm c}\cdot{N\over n}\,
  m_{{\rm v}N}\,m_{{\rm c}N}={n'\over n} m_{{\rm v}} m_{{\rm c}}\,.
   \end{equation}
If Eq.\ (\ref{eq31a}) holds true for a centered system made up of $J$ successive refracting spheres, it holds for a system made up of $J+1$ refracting spheres. Since it holds true for $J=1$, it holds for every $J$.  The proof is complete. The proof also holds if ${\cal D}$ is a spherical mirror.

\subsection{Imaging coherent plane-objects}%***********************************

\subsubsection*{Centered systems with foci} We consider a centred system with foci $F$ and $F'$. Let ${\cal A}$ be a plane object, orthogonal to the optical axis and passing through point $A$. Let ${\cal A}'$ be the coherent image of ${\cal A}$. Where is  ${\cal A}'$ formed and what is its curvature radius? To answer, we apply the double conjugation-law:
  \begin{itemize}
  \item The object ${\cal A}$ passes through $A$: its image ${\cal A}'$ passes through $A'$, the paraxial image of  $A$.
  \item The curvature center of ${\cal A}$ is at infinity: the curvature center of ${\cal A}'$ is at $F'$, which is the paraxial image of a point at infinity (on the optical axis). 
  \end{itemize}

  If $m_a$ denote the lateral magnification between $A$ and $A'$, the field amplitude on  ${\cal A}'$ is
  \begin{equation}
    U_{A'}(x',y')={1\over m_a}\,U_A\left({x'\over m_a},{y'\over m_a}\right).\end{equation}

\begin{figure}[h]%$$$$$$$$$$$$$$$$$$$$$$$$$$$$$$$$$$$$$$$
  \begin{center}
  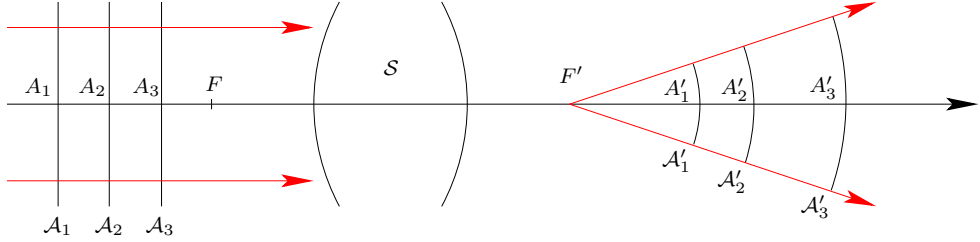
  \caption{\small Coherent imaging of planar objects by a centered system ${\cal S}$ with foci ($F$ and $F'$). Plane ${\cal A}_j$, whose vertex is $A_j$, is  imaged on  the spherical cap ${\cal A}'_j$ passing through $A'_j$ and centered at the image focus $F'$. Point $A'_j$ is the paraxial image of $A_j$. \label{fig5b}}
  \end{center}
\end{figure}

  We note that if ${\cal A}$ is a plane, its coherent image ${\cal A}'$ is properly a sphere, centered on the image focus of the centered system. It is not possible to obtain the coherent image of a plane on a plane, if the system is a centered system with foci. Hence the necessity to consider spherical caps. The result is illustrated in Fig.\ \ref{fig5b} for various planar objects.

  Conversely, the image of a spherical cap centered at the object focus of a lens is a plane.

%$$$$$$$$$$$$$$$$$$$$$$$$$$$$$$$$$$$$$$$$$$$$$

  \subsubsection*{Afocal systems}
\begin{figure}[h]%$$$$$$$$$$$$$$$$$$$$$$$$$$$$$$$$$$$$$$$
  \begin{center}
  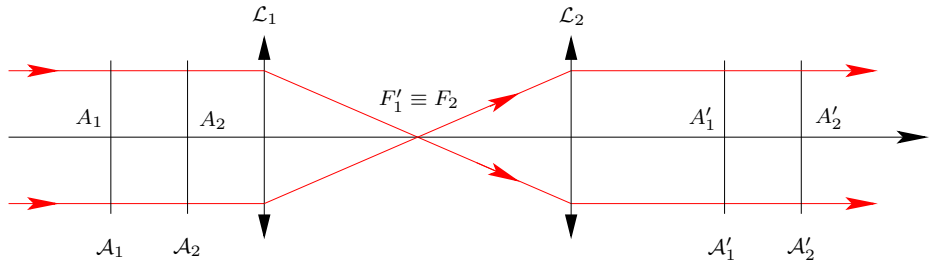
  \caption{\small Coherent imaging of planar objects ${\cal A}_1$ and ${\cal A}_2$ by an afocal  centered system made up of two converging lenses ${\cal L}_1$ and ${\cal L}_2$. The image focus $F'_1$ of ${\cal L}_1$ coincides with $F_2$, the object focus of ${\cal L}_2$. Coherent images ${\cal A}'_1$ and ${\cal A}'_2$ are planar. \label{fig6b}}
  \end{center}
  \vskip -.5cm
\end{figure}

  Let us return to the planar object of the previous paragraph and consider that the system forming its image is afocal (which means that its foci are at infinity). The law of double conjugation thus applies as follows:
   \begin{itemize}
  \item The object ${\cal A}$ passes through $A$: its image ${\cal A}'$ passes through $A'$, the paraxial image of $A$.
  \item The curvature center  ${\cal A}$ lies at infinity: the curvature center of ${\cal A}'$ also lies at infinity, since the optical system is afocal.
   \end{itemize}
   We conclude that the coherent image of a planar object by an afocal system is a plane. The result is illustrated in Fig.\ \ref{fig6b}.

 %$$$$$$$$$$$$$$$$$$$$$$$$$$$$$$$$$$$$$$$$$$$$$

%****************************************************************
\section{Interpreting the radius-magnification law}\label{sect3}
%****************************************************************

\subsection{Radius magnification and longitudinal magnification}\label{sect31}

The notion of radius magnification may be regarded as a generalization of the longitudinal magnification, as used in geometrical optics, and which is as follows. An infinitesimal variation $\D z$ of point $A$ on the optical axis results in an infinitesimal  variation $\D z'$ of the conjugate  $A'$. The longitudinal magnification at $A$ is then defined by $m_z=\D z'/\D z$. From $zz'=ff'$, we deduce: $ z\,\D z'+z'\D z=0$, and then
\begin{equation}
  m_z={\D z'\over \D z}=-{ff'\over z^2}=-{z'^2\over ff'}\,.\label{eq41a}
\end{equation}
Since both $\D z$ and $\D z'$ are infinitesimal quantities, Eq.\ (\ref{eq41a}) makes sense only if $A$ and $A'$ are not too close from the foci of the optical system, because if $A$ is close to $F$, a very small variation $\D z$ of $z$ may result in a large variation of $z'$, and such a variation may not be written $\D z'$.

Since $f'/n'=-f/n$, Eq.\ (\ref{eq41a}) can be written
\begin{equation}
  m_z={n'\over n}\,{z'^2\over f'^2}={n'\over n}\,m^2\,,\label{eq42a}\end{equation}
where $m$ is the lateral magnification for the  conjugation between $A$ and $A'$, given by Eq.\ (\ref{eq3}).

We now show that Eq.\ (\ref{eq42a}) is a limit form of the radius-magnification law of Bonnet.  We consider a spherical cap ${\cal A}$ (vertex $A$, center $C$, radius $R_A$) and its coherent image ${\cal A}'$ (vertex $A'$, center $C'$, radius $R_{A'}$), formed by a centered system. We assume that the radius $R_A$ tends to $0$, that is, the center of curvature $C$ tends to the vertex $A$. Then the center $C'$ tends to $A'$ and $R_{A'}$ tends to $0$. Nervertheless the radius magnification tends to a finite value, because the lateral magnification at centers, $m_{\rm c}$, tends to the lateral magnification at vertices, $m_{\rm v}$. We have
\begin{equation}
  \lim_{R_A\rightarrow \,0}{R_{A'}\over R_A}= \lim_{R_A\rightarrow\, 0}m_{\rm r}= \lim_{C \rightarrow \,A}{n'\over n}m_{\rm v}\,m_{\rm c}={n'\over n}\,m_{\rm v}^2\,.
\end{equation}
If $R_A$ tends to $0$, we write $R_A=\D z$ and $R_{A'}=\D z'$ and we obtain $m_z=m_{\rm v}^2\,n'/n$, which is Eq.\ (\ref{eq42a}), since the lateral magnification at vertices is also the lateral magnification for the conjugation of $A$ and $A'$: $m_{\rm v}=m$.

\begin{figure}[h]%$$$$$$$$$$$$$$$$$$$$$$$$$$$$$$$$$
  \begin{center}
  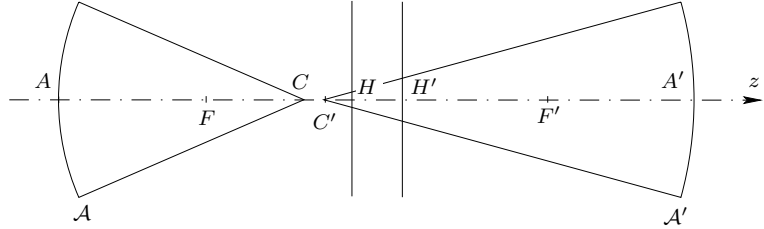
  \caption{\small Illustration of the radius magnification regarded as longitudinal magnification  between longitudinal segments. The optical system is represented by its foci ($F$ and $F'$) and its principal points ($H$ and $H'$). The lateral magnification for $A$ and $A'$ is $m_{\rm a}=-1$; the lateral magnification for $C$ and $C'$ is $m_{\rm c}=3/2$; the radius magnification is $m_{\rm r}=-3/2$,  and is also the magnification between $\overline{A'C'}$ and $\overline{AC}$ (point $C'$ is virtual).\label{fig5}}
  \end{center}
\end{figure}

Since $R_A=\overline{AC}$ and $R_{A'}=\overline{A'C'}$, the radius magnification law appears to be a generalization of Eq. (\ref{eq42a}): it provides the longitudinal magnification for finite distances between two points on the optical axis. The law holds even if $A$ or $C$ are close to the object focus, or lie on both side of the focus. For example in Fig.\ \ref{fig5}, we have: $n=n'$; $f'=-f>0$\,; $\overline{FA}=f$ and $\overline{FA'}=f'$; $\overline{FC}=-2f/3$, and $\overline{F'C'}=-3f'/2$\,;
$m_{\rm v}=m_{\rm a}=-1$, $m_{\rm c}=3/2$, and $m_{\rm r}=-3/2$, that is: $\overline{A'C'}/\overline{AC}=-3/2$. Point $C'$ is virtual.

%$$$$$$$$$$$$$$$$$$$$$$$$$$$$$$$$$$$$$$

\subsection{Radius magnification and longitudinal Lagrange--Helmholtz formula}

\begin{figure}[h]%$$$$$$$$$$$$$$$$$$$$$$$$$$$$$$$$$$$$
  \begin{center}
  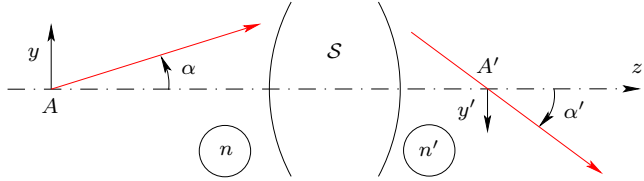
  \caption{\small Lagrange--Helmholtz invariant: $n\,y\,\alpha = n'y'\alpha '$.\label{fig6}}
  \end{center}
\end{figure}

We first recall some laws of paraxial optics. We consider a centered system, as in Fig.\ \ref{fig6}, and a conjugation between a transverse object $y$, located at $A$, and its paraxial image $y'$, located at $A'$; the lateral magnification is $m_y=y'/y$. A ray issued from $A$ makes the angle $\alpha$ with the optical axis and its image is a ray issued from $A'$ and making the angle $\alpha '$ with the axis. The angular magnification is $m_\alpha$ (see Fig.\ \ref{fig6}, where $\alpha >0$ and $\alpha '<0$), defined by
\begin{equation}
  m_\alpha={\alpha '\over\alpha}\,.\end{equation}

The Lagrange--Helmholtz formula \cite{Mar1} (also called Smith--Helmholtz formula \cite{Bor})  is
\begin{equation}
  n'y'\alpha'=n\,y\,\alpha\,\label{eq45a},\end{equation}
and if we denote $m_n=n'/n$, for sake of symmetry,  Eq.\ (\ref{eq45a}) can be written
\begin{equation}
  m_n\,m_y\,m_\alpha =1\,.\end{equation}

%$$$$$$$$$$$$$$$$$$$$$$$$$$$$$$$$$$$$$$$$$$

The longitudinal Lagrange--Helmholtz formula is \cite{Con}
\begin{equation}
  n'{\alpha'}^2\,\D z'=n\,\alpha^2\,\D z\,,\label{eq47}
  \end{equation}
where $\D z$ and $\D z'$ are as in Section \ref{sect31}.  It can also be written
\begin{equation}
  m_n\,m_\alpha^{\; 2}\,m_z=1\,.\end{equation}

We will show that Eq.\ (\ref{eq47}) is a limit form of a more general formula.
We consider the imaging between an object spherical-cap ${\cal A}$ (vertex $A$, center $C$, radius $R_A$) and its coherent image ${\cal A}'$ (vertex $A'$, center $C'$, radius $R_{A'}$). Let $m_\alpha$ be the angular magnification at vertices and $m_\gamma$ the angular magnification at centers. The Lagrange--Helmholtz formula at vertices is
$m_n\,m_{\rm v}\,m_\alpha =1$,  and that at centers is $m_n\,m_{\rm c}\,m_\gamma =1$, so that eventually the radius magnification is
\begin{equation}
  m_{\rm r}={n'\over n}\,m_{\rm v}\,m_{\rm c}=m_n\,m_{\rm v}\,m_{\rm c}=m_n\,{1\over m_n\,m_\alpha}\,{1\over m_n\,m_\gamma}={1\over m_n\,m_\alpha\,m_\gamma}\,,\label{eq47a}
\end{equation}
that is
\begin{equation}
  m_n\,m_{\rm r}\,m_\alpha\,m_\gamma=1\,.\label{eq48a}
\end{equation}
Equation (\ref{eq48a}) can aso be written
\begin{equation}
  n'\,\overline{A'C'}\,\alpha '\gamma '= n\,\overline{AC}\, \alpha \,\gamma \,,\label{eq51}\end{equation}
since $m_{\rm r}=R_{A'}/R_A=\overline{A'C'}/\overline{AC}$.

Il $C$ tends to $A$, then $C'$ tends to $A'$ and $m_\gamma$ tends to $m_\alpha$. Since $R_A$ and $R_{A'}$ tend to $0$, we write $R_A=\D z$ and $R_{A'}=\D z'$, and by Eq.\ (\ref{eq48a}) we obtain
\begin{equation}
  {\D z'\over \D z}={R_{A'}\over R_A}=m_{\rm r}=\lim_{C\rightarrow \,A}{1\over m_n\,m_\alpha\,m_\gamma} 
 ={n\over n'}\,{1\over m_\alpha^{\; 2}}={n\over n'}\,{\alpha ^2\over {\alpha '}^2}\,, \end{equation}
which is Eq.\ (\ref{eq47}) once more.

\begin{figure}[h]%$$$$$$$$$$$$$$$$$$$$$$$$$$$$$$$$$
  \begin{center}
  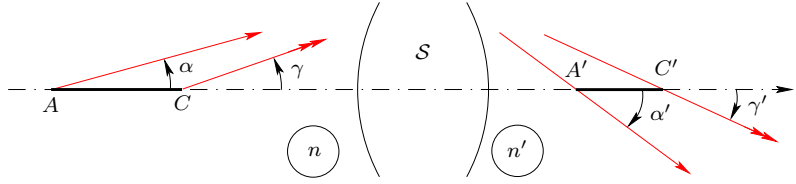
  \caption{\small Longitudinal Lagrange--Helmholtz invariant for finite segments: $n\,\overline{AC}\,\alpha\,\gamma = n'\,\overline{A'C'}\alpha '\gamma '$.\label{fig7}}
  \end{center}
\end{figure}

Equation (\ref{eq51}) is a generalization of the longitudinal Lagrange--Helmholtz formula that holds for finite longitudinal segments \cite{Det} (see Fig. \ref{fig7}).

%$$$$$$$$$$$$$$$$$$$$$$$$$$$$$$$$$$$$$

\begin{figure}[h]%$$$$$$$$$$$$$$$$$$$$$$$$$$$$$$$$$
  \vskip -.2cm
  \begin{center}
  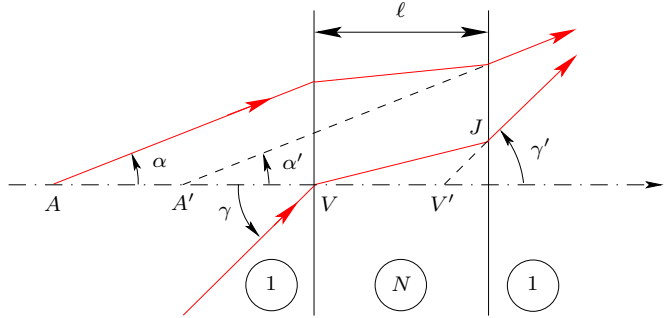
  \caption{\small Imaging through a plane parallel plate. Since $\alpha '=\alpha$ and  $\gamma '=\gamma$, the generalized longitudinal Lagrange--Helmholtz formula leads to $\overline{AA'}=\overline{VV'}$.\label{fig7b}}
  \end{center}
\end{figure}

We illustrate the generalized longitudinal Lagrange--Helmholtz by an example borrowed from Dettwiller \cite{Det}.
We consider a plane parallel plate of width $\ell$, refractive index $N$, and placed in the air. (The plate is an afocal system.) Let $V$ be a point on  the plate (Fig.\ \ref{fig7b}): its image  is $V'$.  Let $A$ and $A'$ be conjugates, with $AV$ orthogonal to the plate. We have $\alpha '=\alpha $, $\gamma '=\gamma$ (see Fig.\ \ref{fig7b});  we set $n=n'=1$, $C=V$ and $C'=V'$ in  Eq.\ (\ref{eq51}), and we obtain  
$\overline{AV}=\overline{A'V'}$. Then $\overline{AA'}=\overline{AV}+\overline{VV'}+\overline{V'A'}=\overline{VV'}$. Kepler's law, applied at point $J$, leads to $\overline{VV'}=(N-1)\ell/N$, so that for every pair of conjugate points $A$ and $A'$ we obtain the  well-known formula
\begin{equation}
  \overline{AA'}={(N-1)\ell\over N}\,.\label{eq52}\end{equation}

%$$$$$$$$$$$$$$$$$$$$$$$$$$$$$$$$$$
  
%******************************************************
  \section{Application to Gaussian beams }\label{sect4}
%******************************************************

\subsection{Gaussian beams}%****************************

The existence of Gaussian beams and their properties are usually developed  in physical optics, in general in the framework of a scalar theory of diffraction \cite{Kog,PPF1} or from the Helmholtz wave-equation \cite{Yar},  not in the scope of geometrical optics.
We will  show, however, that  imaging Gaussian beams can be explained very well in the framework of geometrical optics, provided that the properties of coherent imaging are applied to it. We begin %, in this section,
by characterizing Gaussian beams.

\goodbreak
Gaussian beams, which are coherent beams, are mainly encountered in the following cases:
\begin{itemize}
    \item In quadratic index media \cite{Yar}.
  
\item In stable laser-cavities and at the output of these lasers \cite{Sie}. %\footnote{This is an approximation that depends on the Fresnel number of the laser cavity. For most lasers this is a very good approximation.}.
 
     \item As an approximation of optical fields in fiber optics \cite{Sny}.  The fundamental mode of an optical fiber (HE$_{11}$ mode) can be approximated by a Gaussian function.
   \end{itemize}

The beam emitted by a laser is usually called a Gaussian beam. The amplitude of the field on a wavefront of such a beam is represented by a Hermite-Gauss function, in the form
\begin{equation}
    U(x,y)=U_0\,H_i(x)\,H_j(y) \,\exp\left(-{x^2+y^2\over w^2}\right)\,,\label{eq31}\end{equation}
where $U_0$  is a  dimensional constant and $H_i$ the Hermite polynomial of order  $i$ ($i$ a natural integer).

Here are the properties of Gaussian beams that will allow us to study their imagings by centered optical systems \cite{PPF1,Chi}:  \newcounter{listado}
  \begin{list}{(\roman{listado})}{\usecounter{listado}}
  \item Wavefronts of a given Gaussian beam are approximated, up to second order, by spherical caps, not concentric, but all centred on the beam axis. The field amplitudes on those surfaces take the form given by Eq.\ (\ref{eq31}), where $w$ depends on the considered wavefront.
  \item Parameter  $w$ has a minimum value, denoted $w_0$, and the corresponding wavefront is a plane, denoted ${\cal W}_0$. In this plane, the area for which the irradiance is greater or equal to $1/\E^2$ is called the waist of the Gaussian beam. The parameter $w_0$ is the (transverse) radius of the waist.
  \item The curvature radius of the wavefront whose vertex is at a distance $d$ ($d$ is an algebraic measure) from the waist is 
    \begin{equation}
      R_d=-d-{\pi^2{w_0}^4\over \lambda^2d}\,,\label{eq53}\end{equation}
   and the transverse radius of the beam on this spherical cap is $w_d$ such that
    \begin{equation}
      {w_d}^2={w_0}^2+{\lambda^2d^2\over \pi^2{w_0}^2}\,,\label{eq54}\end{equation}
where $\lambda$ is the wavelength of radiations.
\item \label{prop4}Given $\lambda$ and given a spherical cap,   there is a unique Gaussian beam for which the transverse radius takes a given value $w$ on the cap. If $R$ denotes the curvature radius of the spherical cap, the transverse radius of the beam waist is given by  
  \begin{equation}
    {w_0}^2={w^2\over 1+\displaystyle{\pi^2w^4\over \lambda^2R^2}}\,,\end{equation}
 and the distance from the plane of the waist to the vertex of the spherical cap is 
  \begin{equation}
    d=-{R\over 1+\left(\displaystyle{\lambda R\over \pi w^2}\right)^2}\,.\end{equation}
    \end{list}

 \subsection{Imaging a Gaussian beam  by a centered system with foci.}

 \subsubsection*{The waist image is not the image waist (the image of the beam-waist  is not the waist of the image-beam)}
 
 Let us consider a Gaussian beam incident on a centered system ${\cal S}$ with foci, and  let ${\cal W}$ be  a wavefront (spherical cap) of the beam. The centered system transforms ${\cal W}$ into a spherical cap ${\cal W}'\!$, and the amplitude of the field on ${\cal W}'$ is Gaussian (it is a homothetic copy of the field on ${\cal W}$). Property (iv)  above allows us to conclude that the incident Gaussian beam is transformed into a Gaussian beam (in the image space).

 Let us highlight an important property of imaging Gaussian beams by optical systems with foci. Let ${\cal W}_0$ be the waist plane of the incident beam, assumed to  be located at  the point $W_0$ of the optical axis (in the object space). The image of $W_0$ is $W'_0$, a point on the optical axis (in the image space). Since it is a plane,  ${\cal W}_0$ is centered at infinity and its coherent image  ${\cal W}'_0$  is properly spherical: it is the spherical cap that passes through $W'_0$ and that is centered at the image focus $F'$ of the centered system (a consequence of curvature center conjugation). This sphere cannot contain the waist of the image beam, because the waist should be on a plane:  the waist of the object beam and the waist of the image beam are not conjugates. There is therefore not imaging between the waists!

\subsubsection*{Pseudo-conjugation formula}

Waists are not conjugated. Then where  is the waist of the image beam located? It has to be on a plane, that is, on a ``sphere'' centered at infinity (in the image space). That plane has then to be the image of the spherical cap that, among all the wavefronts of the incident beam, is centered at the object focus $F$ of the centered system, because $F$ is the unique point (on the optical axis, in the object space) whose image is precisely at infinity. All that is a consequence of the curvature-center conjugation law (a part of the double-conjugation property).
Let us show that such a spherical cap indeed exists.

 \begin{figure}[b]%$$$$$$$$$$$$$$$$$$$$$$$$$$$$$$$$$
  \begin{center}
  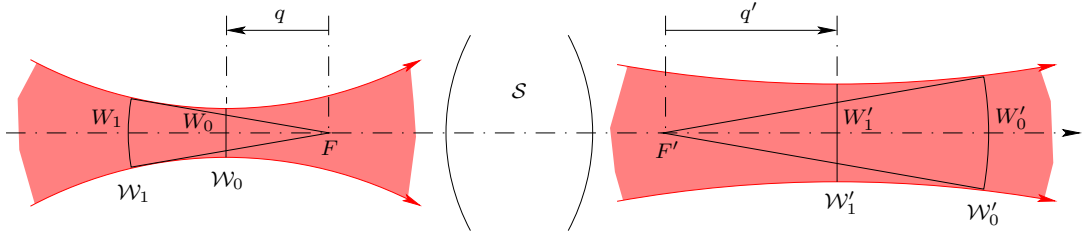
  \caption{\small Imaging a Gaussian beam by a centered system with foci.\label{fig8}}
  \end{center}
\end{figure}%$$$$$$$$$$$$$$$$$$$$$$$$$$$$$$$$$$$$$$$

We denote $q=\overline{FW_0}$ and look for the spherical wavefront ${\cal W}_1$ (of the object beam) that is centred at $F$; its vertex is denoted $W_1$ (see Fig.\ \ref{fig8}). If $d_1=\overline{W_0W_1}$, the radius of ${\cal W}_1$ is $R_1$ such that $R_1=\overline{W_1F}$, and Eq.\ (\ref{eq53}) leads to
\begin{equation}
   q+d_1=\overline{FW_1}=-R_1=d_1+{\pi^2{w_0}^4\over \lambda^2d_1}\,,\end{equation}
so that
 \begin{equation}
   d_1={\pi^2{w_0}^4\over \lambda^2q}\,.\label{eq37}\end{equation}
 The spherical cap ${\cal W}_1$, which is a wavefront of the incident Gaussian-beam,  is then totally defined: its vertex is $W_1$, at a distance $d_1$ of $W_0$, its center of curvature is the object focus $F$. Since ${\cal W}_1$ is centered at $F$, its image is centered at infinity and is a plane, denoted ${\cal W}_1'$. That plane is a wavefront of the image beam,  because it is the coherent image of a wavefront of the object beam, and since the waist of a Gaussian beam is on a plane, the  plane ${\cal W}'_1$ contains the waist of the image beam.
 
If $W'_1$ denotes the image of $W_1$, and  if $q'=\overline{F'W'_1}$, the Newton's conjugation-formula gives
 \begin{equation}
   ff'=\overline{FW_1}\cdot \overline{F'W'_1}=(q+d_1)q'=\left(q+{\pi^2{w_0}^4\over \lambda^2q}\right)q'\,,
   \end{equation}
 that  is
 \begin{equation}
   q'={ff'\over q+\displaystyle{\pi^2{w_0}^4\over \lambda^2q}}\,,\label{eq60}\end{equation}
which is often written in the form
  \begin{equation}
    qq'={ff'\over 1+\displaystyle{\pi^2{w_0}^4\over \lambda^2q^2}}\,.\label{eq61}\end{equation}

  Equation(\ref{eq61})---or (\ref{eq60})---gives the position of the waist of the image beam when the position of the waist of the object beam is known. In particular Eq.\ (\ref{eq61}) resembles Newton's conjugation formula ($zz'=ff'$), with origins at foci: it is therefore called usually ``waist conjugation-formula'', which is abusive, since there is no conjugation of waists, as explained above.

\subsubsection*{Image-beam waist dimension}

Let $w_1$ be the transverse radius of the beam on the previous spherical cap ${\cal W}_1$ and let $m$ be the lateral magnification of the conjugation of $W_1$ with $W'_1$. The waist radius of the image beam is $w'_1=mw_1$, and
\begin{equation}
    m=-{f\over \overline{FW_1}}={f\over R_1}\,.
  \end{equation}
  Equations (\ref{eq53}) and (\ref{eq54}) lead us to write
  \begin{eqnarray}
    {w'_1}^2={f^2{w_1}^2\over {R_1}^2}={f^2{w_0}^2\left(1+\displaystyle{\lambda^2{d_1}^2\over \pi^2{w_0}^4}\right)\over {d_1}^2\left(1+\displaystyle{\pi^2{w_0}^4\over \lambda^2 {d_1}^2}\right)^2}\!\!\!
    &=&\!\!\!{f^2\lambda^2\over \pi^2{w_0}^2}\cdot{1\over 1+\displaystyle{\pi^2{w_0}^4\over \lambda^2{d_1}^2}}
    \nonumber \\
    &=&\!\!\!{f^2\lambda^2\over \pi^2{w_0}^2}\cdot{1\over 1+\displaystyle{\lambda^2q^2\over \pi^2{w_0}^4}}\,,\label{eq63}\end{eqnarray}
  where the last equality results from Eq.\ (\ref{eq37}).

\subsubsection*{Some examples}
Here are some illustrations of the previous results.
 \begin{itemize}
 \item We assume $q=0$, which means that the waist of the object beam is in the object focal-plane of the centered system. By Eq.\ (\ref{eq60}) we obtain $q'=0$: the waist of the image beam is in the image focal-plane of the centered system. Clearly there is no conjugation of waists! The waist radius of the image beam is
    \begin{equation}
      w'_1={|f|\lambda \over \pi w_0}\,.\end{equation}
  \item We assume  $|q|>>{w_0}^2/\lambda$. Equation (\ref{eq61}) is approximated by  $qq'=ff'$ (Newton's conjugation-formula, which appears as a limit case). The radius of the waist of the image beam ($w'_1$) then tends to $0$.
  \item  In the same vein, consider the limit case $w_0=0$ (with $q\ne 0$). Equation (\ref{eq61}) then reduces to Newton's conjugation-formula: $qq'=ff'$. According to Eq.\ (\ref{eq63}), the transverse radius $w'_1$ tends to $0$: both waists reduce to points and their ``conjugation law'' tends to the usual  conjugation law of paraxial geometrical optics.
     \end{itemize}

\subsection{Imaging a Gaussian beam by an afocal system}%**********************
%It is known that
An afocal system forms a planar coherent-image of a planar object. If a Gaussian beam is incident on an afocal system, the waist having to be on a plane, the waist of the image beam is the image of the waist of the incident beam (object beam). There is conjugation of the waists. If the waist ${\cal W}_0$ of the incident beam is at $W_0$, its coherent image ${\cal W}'_0$ through a given afocal system passes through $W'_0$, the paraxial image of $W_0$. Since ${\cal W}_0$  is centered at infinity,  ${\cal W}'_0$ also is centered at infinity, because  the ``image focus'' of the afocal system is at infinity:  ${\cal W}'_0$  is a plane and it is on this plane that the waist of the image beam is located.

\begin{figure}[h]%$$$$$$$$$$$$$$$$$$$$$$$$$$$$$$$$$
  \begin{center}
  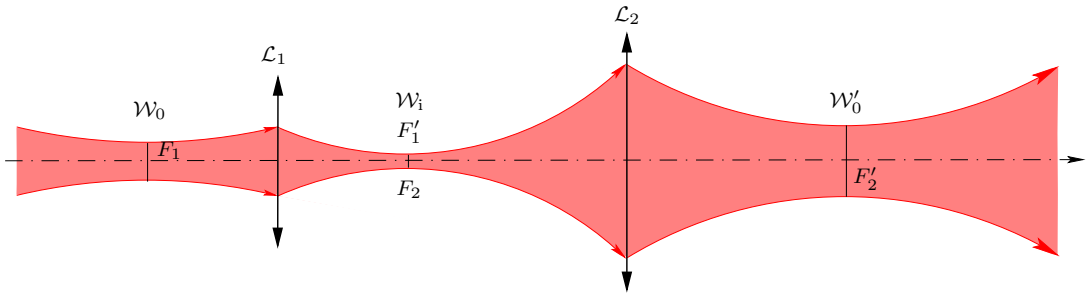
\caption{\small Imaging a Gaussian beam by an afocal centered system made up of two convergent lenses.\label{fig9}}
  \end{center}
\end{figure}

Let us illustrate that with an example. Consider an afocal system ${\cal L}$ consisting of two convergent thin lenses ${\cal L}_1$ and ${\cal L}_2$, with respective image focal-lengths $f'_1$ and $f'_2 $. The object focus $F_2$ of ${\cal L}_2$ coincides with the image focus $F'_1$ of ${\cal L}_1$ (see Fig.\ \ref{fig9}). The image of the object focus $F_1$ of ${\cal L}_1$ through ${\cal L}$ is the image focus $F'_2$ of ${\cal L}_2$. Consider a Gaussian beam  with waist ${\cal W}_0$ placed at $F_1$, the transverse radius of the waist being $w_0$. The first example at the end of the previous section shows that the waist ${\cal W}_{\rm i}$ of the intermediate beam, which is the image beam through ${\cal L}_1$, lies in the image focal-plane of ${\cal L}_1$ and its transverse radius is $w_{\rm i}= \lambda f'_1/(\pi w_0)$. Since $F_2$ merges with $F'_1$, the same property applies to imaging through ${\cal L}_2$: the waist of the image through ${\cal L}_2$ of the previous intermediate beam is in the focal plane of ${\cal L}_2$ and its transverse radius is
 \begin{equation}
    w'_0={\lambda f'_2\over \pi w_{\rm i}}={f'_2\over f'_1}w_0=m\,w_0\,,\end{equation}
where $m$ denotes the lateral magnification of the afocal system (which is a constant, independant of the considered conjugation).
There is imaging between the Gaussian-beam incident on ${\cal L}_1$ and the Gaussian-beam emerging from ${\cal L}_2$, according to the laws of paraxial geometrical optics.

%$$$$$$$$$$$$$$$$$$$$$$$$$$$$$$$$$$$$$

%*******************************
\section{Conclusion}\label{conc}
%*******************************

Coherent imaging of spherical caps by centered systems, that is imaging both amplitudes and phases, is obtained under two conditions: (i) vertex conjugation, in accordance with paraxial geometrical optics; (ii)~curvature center conjugation, which is properly a coherent condition for phase preserving and which  requires a second-order approximation.  These two conditions constitute the ``double-conjugation law'' of coherent imaging and they have been deduced from pure methods of geome\-trical optics.  To paraxial geometrical optics is therefore added coherent geometrical optics, or metaxial geometrical optics, whose core is the double-conjugation law. Imaging Gaussian beams by centered systems falls within this framework.

%************************************ REFERENCES*****************************

\end{document}